\documentclass[ALICE,manyauthors]{cernphprep}

\usepackage{hyperref}

\begin{document}
\def\pt{$p_{\mathrm{T}}$}
\def\sqs{$\sqrt{s}$~}
\def\sqsnn{$\sqrt{s_{\rm{NN}}}$~}
\def\dzero{$\ensuremath{d_{0}}$}
\hyphenation{analy-ses}
\hyphenation{analy-sis}

\begin{titlepage}

\PHyear{2014}
\PHnumber{089}              
\PHdate{09 May}              

\title{Beauty production in pp collisions at $\mathbf{\sqrt{s} =}$ 2.76 TeV \protect \\ measured via semi-electronic decays}
\ShortTitle{Beauty production in pp collisions at $\sqrt{s} =$ 2.76 TeV}  

\Collaboration{ALICE Collaboration%
        \thanks{See Appendix~\ref{app:collab} for the list of collaboration
                      members}}
\ShortAuthor{ALICE Collaboration}      

\begin{abstract}
The ALICE collaboration at the LHC reports measurement of the inclusive production cross section of electrons from semi-leptonic decays of beauty hadrons with rapidity $|y|<0.8$ and transverse momentum $1<p_{\mathrm{T}}<10$ GeV/$c$, in pp collisions at $\sqrt{s} = $ 2.76 TeV. Electrons not originating from semi-electronic decay of beauty hadrons are suppressed using the impact parameter of the corresponding tracks. The production cross section of beauty decay electrons is compared to the result obtained with an alternative method which uses the distribution of the azimuthal angle between heavy-flavour decay electrons and charged hadrons. Perturbative QCD predictions agree with the measured cross section within the experimental and theoretical uncertainties. The integrated visible cross section, $\sigma_{\mathrm{b} \rightarrow \mathrm{e}} = 3.47\pm0.40(\mathrm{stat})^{+1.12}_{-1.33}(\mathrm{sys})\pm0.07(\mathrm{norm})~ \mu$b, was extrapolated to full phase space using Fixed Order plus Next-to-Leading Log (FONLL) calculations to obtain the total b$\bar{\mathrm{b}}$ production cross section, $\sigma_{\mathrm{b\bar{b}}} = 130\pm15.1(\mathrm{stat})^{+42.1}_{-49.8}(\mathrm{sys})^{+3.4}_{-3.1}(\mathrm{extr})\pm2.5(\mathrm{norm})\pm4.4(\mathrm{BR})~ \mu$b.
\end{abstract}
\end{titlepage}
\setcounter{page}{2}

\section{Introduction}
\label{introduction}

Perturbative Quantum Chromodynamics (pQCD) calculations of the production of heavy (charm and beauty) quarks can be carried out with well-controlled accuracy, due to the hard (high $Q^2$) scale imposed by the large mass of heavy quarks \cite{fonll3,gmvfns,Maciula:2013wg}. In addition, the large mass implies that heavy quark production in high energy collisions of heavy ions occurs early compared to the formation time of the strongly interacting partonic matter generated in such collisions \cite{brahmsqgp,phenixqgp,phobosqgp,starqgp}. Therefore, the study of heavy quark production in pp collisions is of interest for two reasons: the measurement of their production cross section provides essential tests of pQCD, and such measurements yield the necessary reference for the corresponding measurements performed in heavy-ion collisions. Properties of the strongly interacting, partonic medium generated in high energy heavy-ion collisions are studied using various heavy-quark observables \cite{DmesonRaa,Dmesonv2,cmsbjet,cms-nonpromptjpsi}.

The objective of the analyses presented here is to obtain the total beauty production cross section by measuring the \pt -differential inclusive production cross section of electrons from semi-electronic decays of beauty hadrons. The measurement is performed in the mid-rapidity region ($|y| < $ 0.8) with the ALICE detector for $1 < p_{\mathrm T} < 10$ GeV/$c$, in pp collisions at \sqs = 2.76 TeV. The total b$\mathrm{\bar{b}}$ production cross section is determined by the extrapolation of the measured \pt -differential production cross section to  full \pt\ and $y$ ranges. The measured relative beauty contribution to the heavy-flavour decay electrons and the inclusive production cross section of electrons from semi-electronic decays of beauty hadrons are compared to the predictions from three different pQCD calculations (FONLL \cite{fonll3}, GM-VFNS \cite{Bolzoni:2012kx}, and $k_{\rm T}$-factorization \cite{Maciula:2013wg}). The primary analysis presented here uses a track impact parameter discriminant, which takes advantage of the relatively long lifetime of beauty hadrons ($c\tau \sim 500~ \mu$m) compared to charm hadrons. A second method discriminates beauty from charm production using the distribution of the azimuthal angle between heavy-flavour decay electrons and charged hadrons, $\Delta\varphi$. For beauty hadron decays the width of the near-side peak, $\Delta\varphi$ around zero, is indeed larger than that of charm hadron decays, due to the decay kinematics of the heavier mass beauty hadrons. The difference is exploited to measure the relative beauty contribution to the heavy-flavour decay electron population, which can be used along with the measured heavy-flavour electron spectrum to compute the production cross section of electrons from beauty hadron decays. 

\section{Event and track selection}
\label{exp}

The data set used for these analyses was recorded during the 2011 LHC run with pp collisions at \sqs = 2.76 TeV. The Minimum Bias (MB) collisions were triggered using the V0 scintillator detectors, located in the forward (2.8 $< \eta < $ 5.1) and backward (-3.7 $< \eta < $ -1.7) regions, and the Silicon Pixel Detector (SPD), which is the innermost part of the Inner Tracking System (ITS). The SPD consists of two cylindrical layers of hybrid silicon pixel assemblies, covering a pseudo-rapidity interval $|\eta| < $ 2.0 and $|\eta| < 1.4 $ for the inner and outer layer, respectively. Both the V0 and SPD detectors cover the full azimuth. The MB trigger required at least one hit in either of the V0 scintillator detectors or in the SPD, in coincidence with the presence of an LHC bunch crossing. Additional details can be found in \cite{alice_charm_xsec}. The MB trigger cross section was measured to be 55.4$\pm$1.0 mb using a van der Meer scan \cite{alice_cross_section}. A fraction of MB events were triggered independently of the read-out state of the Silicon Drift Detector (SDD), which equips the two intermediate layers of the ITS. The Electromagnetic Calorimeter (EMCal) is a sampling calorimeter based on Shashlik technology, covering a pseudo-rapidity interval $|\eta| < $ 0.7 and covering $100^{\circ}$ in azimuth \cite{rongrong}. The EMCal Single Shower (SSh) trigger system generates a fast energy sum (800 ns) at Trigger Level 0 for overlapping groups of 4$\times$4 ($\eta \times \varphi$) adjacent EMCal towers, followed by comparison to a threshold energy \cite{Kral:2012ae}. The data set recorded with the EMCal trigger required that the MB trigger condition was fulfilled, and that at least one SSh sum exceeded a nominal threshold energy of 3.0 GeV. The results reported are based on 51.5 million MB events (integrated luminosity of 0.9 nb$^{-1}$)  and 0.64 million EMCal triggered events (integrated luminosity of 14.9 nb$^{-1}$). The impact parameter analysis was performed solely on the MB sample. The method based on the distribution of the azimuthal angle between heavy-flavour decay electrons and charged hadrons (i.e. electron-hadron correlation) was done using both the MB and EMCal trigger samples. In the offline analysis, events which satisfied the trigger conditions were required to have a collision vertex with at least two tracks pointing to it and the vertex position along the beam line to be within $\pm$10 cm of the nominal center of the ALICE detector. 

Charged particle tracks were reconstructed offline using the Time Projection Chamber (TPC) \cite{tpc} and the ITS \cite{its}. To have a homogeneously reconstructed sample of tracks, the SDD points were always excluded from the track reconstruction used for these analyses. EMCal clusters were generated offline via an algorithm that combines signals from adjacent EMCal towers. The cluster size was constrained by the requirement that each cluster contains only one local energy maximum. In the case of the EMCal-based analysis, charged tracks were propagated to the EMCal and matched to clusters in the EMCal detector. The matching required the difference between the cluster position and track extrapolation at the EMCal surface to be smaller than 0.025 units in $\eta$ and 0.05 radians in $\varphi$. 

\begin{figure}[tbh!]
\centering
\includegraphics[width=0.9\textwidth]{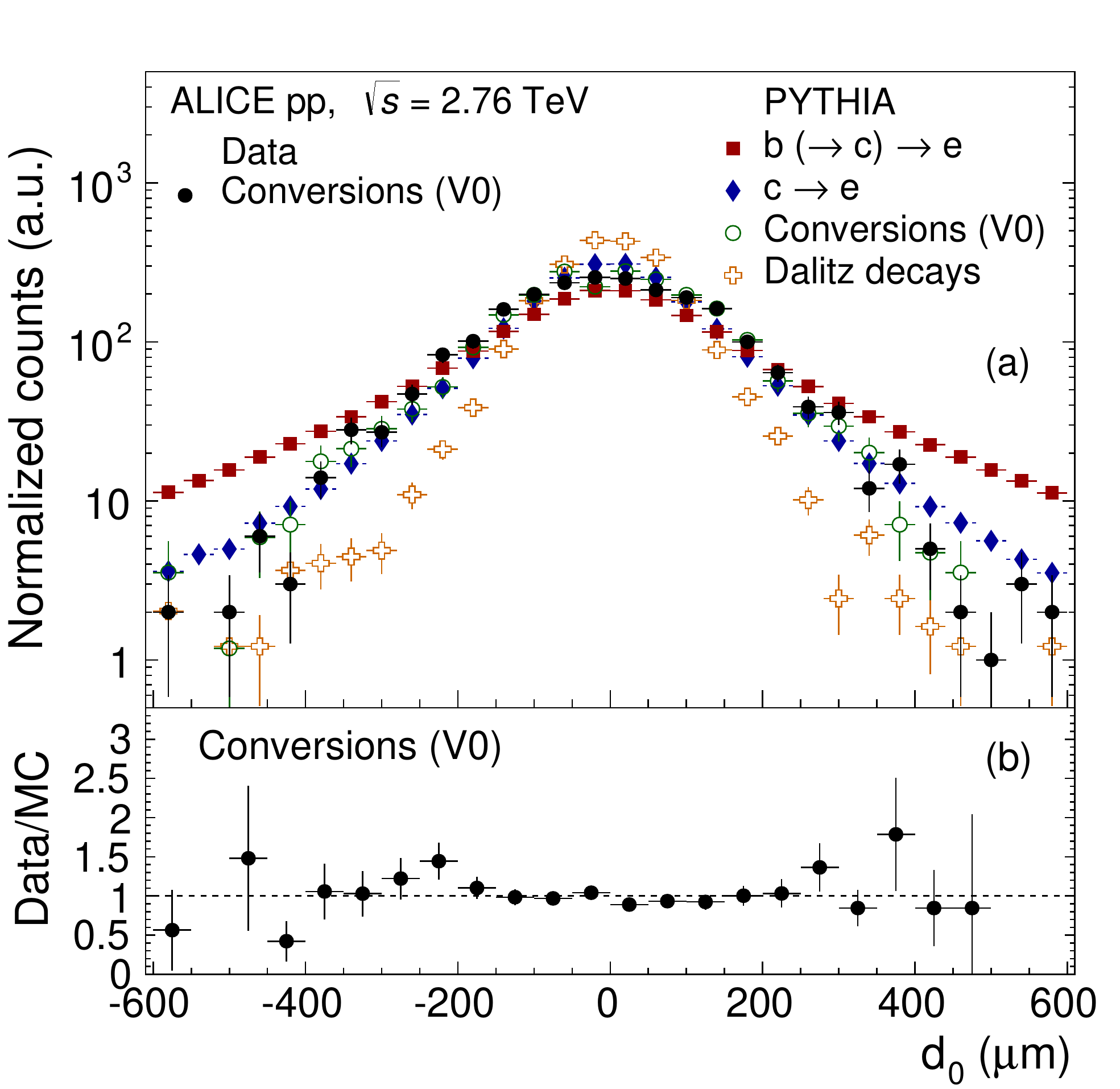}
\caption{(Color online) (a) Transverse impact parameter ($d_{0}$) distributions of electrons from beauty and charm hadron decays, light hadron decays, and photon conversions obtained with PYTHIA  6 simulations in the electron \pt\ range 1 $<$ \pt\ $<$ 6 GeV/$c$, along with the measured distribution of conversion electrons. The distributions are normalized to the same integrated yield. (b) Ratios of the measured and simulated $d_{0}$ distributions of conversion electrons in the ranges 1 $<$ \pt\ $< $ 6 GeV/$c$.}
\label{fig:impactpara}       
\end{figure}

Electrons were identified using the TPC, Time of Flight (TOF), and EMCal detectors \cite{alice_hfe}. Background hadrons, in particular charged pions, were rejected using the specific energy loss, d$E$/d$x$, of charged particles measured in the TPC. Tracks were required to have a d$E$/d$x$ value between one standard deviation below and three standard deviations above the expected value for electrons. In the low momentum region (below 2.0 GeV/$c$ for the impact parameter analysis and below 2.5 GeV/$c$ for the correlation analysis) electron candidates were required to be consistent within three standard deviations with the electron time of flight hypothesis. TOF-based discrimination is not efficient at higher transverse momentum and the TOF was not required. The EMCal-based correlation analysis required $E/p$ to be within a window of 0.8 and 1.2 times the nominal value of $E/p$ for electrons, where $E$ is the energy deposited in the EMCal and $p$ is the track momentum measured in the tracking system. Tracks were required to have hits in the SPD in order to suppress the contribution of electrons that originated from photon conversions in the inner tracking detector material and to improve the resolution on the track impact parameter. 

\section{Analysis}
\label{meth}

\subsection{Impact parameter technique}

The measured electron sample contains contributions from beauty and charm hadron decays, along with background sources. The background is primarily composed of electrons from photon conversions in the beam-pipe and ITS material, $\pi^0$ and $\eta$ Dalitz decays, and di-electron decays of light neutral vector mesons. The relative contribution of electrons from beauty hadron decays can be enhanced by selecting on the displacement of electron tracks from the primary vertex of the pp collision, as described in detail in \cite{beauty7pp}.  

The relatively long lifetime of beauty hadrons was exploited by selecting on the transverse impact parameter (\dzero), which is the projection of the charged track distance of closest approach to the primary vertex vector onto the transverse plane, perpendicular to the beamline. The sign of \dzero\ is given according to the track position relative to the primary vertex after the track has been spatially extended in the direction perpendicular to its \pt\ vector. The resolution of \dzero\, is better than 85 $\mu$m for $p_{\mathrm T}  > 1 \ \mathrm{GeV}/c$. Fig. \ref{fig:impactpara} (a) shows the impact parameter distribution for all significant contributions to the measured electron sample in the range $1 < p_{\mathrm T}  < 6 \ \mathrm{GeV}/c$. The distributions were obtained using a Monte Carlo (MC) simulation with GEANT3 \cite{geant}, where the pp collisions were produced using the PYTHIA 6 event generator (Perugia-0 tune) \cite{perugia0}. Each source has a distinct \dzero\ distribution. The \dzero\ distribution of electrons from Dalitz decays is relatively narrow compared to that from beauty hadron decays, since Dalitz electrons are effectively generated at the collision vertex. The charm hadron decay and conversion electron \dzero\ distributions are broader than that of the Dalitz decay distribution since they emerge from secondary vertices, but are not as broad as those from beauty decays. For comparison, the \dzero\ distribution of conversion electrons from data is also shown in the figure. This pure sample of electrons from photon conversions in the detector material was identified using a V0-finder and an optimized set of topological selection requirements. Fig. \ref{fig:impactpara} (b) shows the ratio of the impact parameter distribution from data to that from simulation in the range $1 < p_{\mathrm T}  < 6 \ \mathrm{GeV}/c$. The ratio is close to unity, showing good agreement of the simulation and measurement of photon conversion electron candidates.

\begin{figure}[tbh!]
\centering
\includegraphics[width=0.9\textwidth]{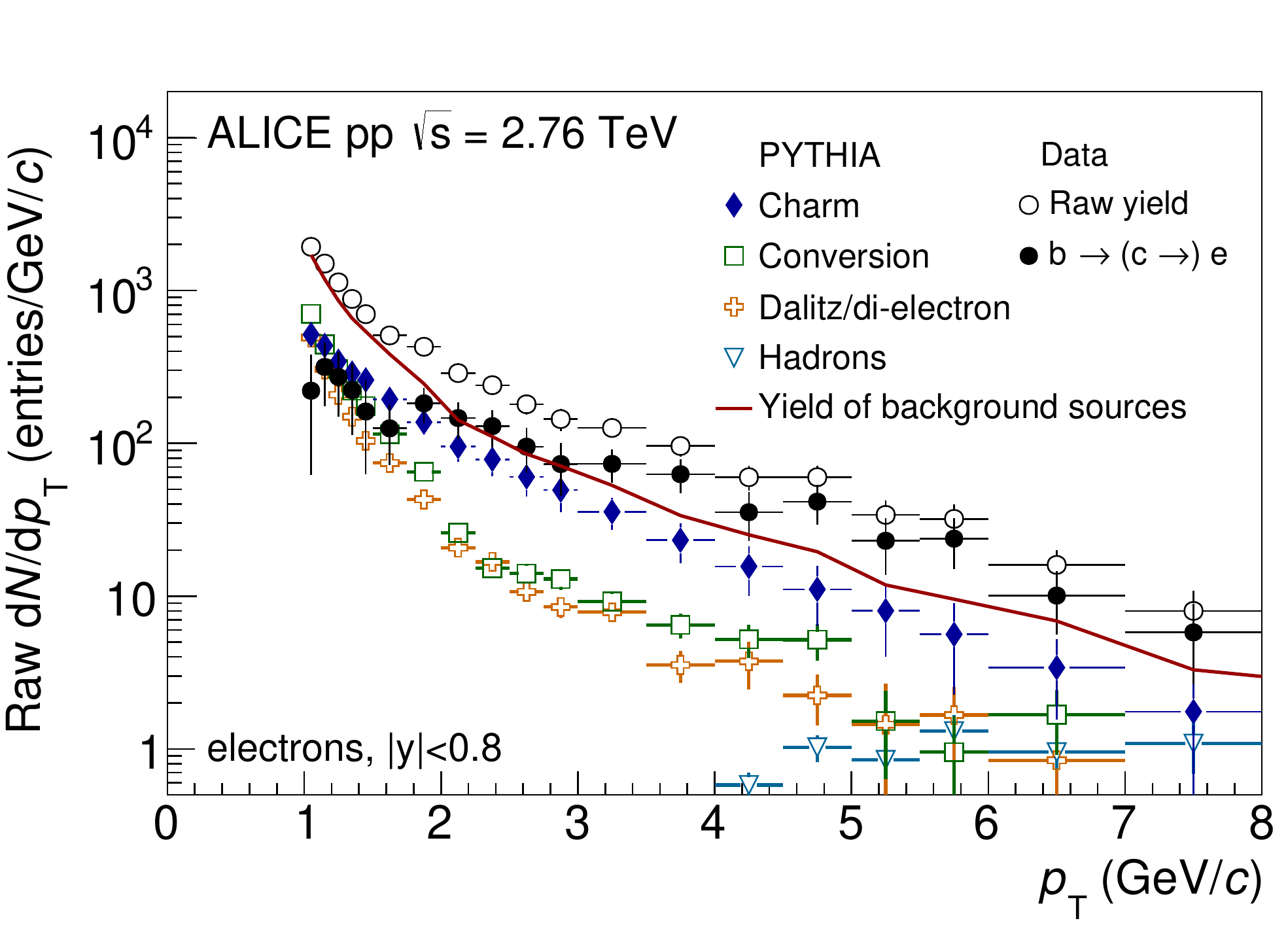}
\caption{(Color online) Raw spectrum of electrons from the impact parameter analysis (open circles) compared to background sources (from charm hadron decays, photon conversions, Dalitz decays, and hadron contamination) as a function of \pt. The background sources originating from light flavour hadrons were obtained using a MC simulation and reweighted according to the $\pi^{0}$ \pt\ spectrum measured with ALICE \cite{alice_pieta}. The charm hadron decay background was estimated using the charm hadron spectra measured with ALICE \cite{alice_d2h}. The raw yield after background sources are subtracted is also shown (filled circles). The error bars represent the statistical uncertainties.}
\label{fig:rawpt}       
\end{figure}

A selection on the transverse impact parameter \dzero\ was applied in order to maximize the signal to background (S/B) ratio of electrons from beauty hadron decays. The requirement on the minimum impact parameter is \pt\ dependent, since the width of the \dzero\ distribution depends on \pt. The S/B ratio varies with \pt\ due to different impact parameter selection efficiency for the various sources. Therefore, separate \pt -dependent parameterizations of the \dzero\ selection requirement were obtained for the analyses which utilize TPC-TOF and TPC-only for electron selection. Electron candidates accepted for the TPC-TOF analysis satisfied the condition $|$\dzero$|$ $>$ 64 + 480$\cdot$exp(-0.56 \pt) (with \dzero\ in $\mu$m and \pt\ in GeV/$c$), while $|$\dzero$|$ $>$ 54 + 780$\cdot$exp(-0.56 \pt) was required for the TPC-only analysis. 

The raw \pt\ distribution of electrons, after the application of track selection criteria, is shown in Fig. \ref{fig:rawpt}, along with the \pt\ distributions of electrons from the various background sources (charm hadron decays, photon conversions, Dalitz/di-electron decays, and hadron contamination). The background distributions were obtained from a MC simulation, with GEANT3. The \pt\ distributions of the background sources were normalized to the total number of events which passed the event selection requirements, and were corrected for the efficiency to reconstruct a primary collision vertex. Among all background contributions, Dalitz decay electrons and photon conversions are dominant at low \pt, where more than 80$\%$ of the background can be attributed to $\pi^{0}$ Dalitz decays and conversions of photons from $\pi^{0}$ decays. At high \pt\ the contribution from charm hadron decays is significant. The contribution from heavy quarkonia decays also becomes significant at high \pt, although this contribution is strongly suppressed in the analysis since the selection on \dzero\ strongly suppresses tracks from such decays. The PYTHIA  simulation does not precisely reproduce the \pt -differential spectra of background sources measured in data. Therefore, the sources of background electrons simulated with PYTHIA were reweighted according to the $\pi^{0}$ \pt\ spectrum measured with ALICE \cite{alice_pieta} and were then propagated in the ALICE apparatus using GEANT3. The spectra of other light mesons were estimated via $m_{\mathrm{T}}$-scaling of the $\pi^{0}$ spectrum. The electron background from charm hadron decays was estimated based on the charm hadron spectra measured with ALICE. The D meson production cross sections were obtained by applying a \sqs scaling to the cross sections measured at  \sqs = 7 TeV~\cite{alice_d2h}. The scaling factor was defined as the ratio of the cross sections from the FONLL calculations at 2.76 and 7 TeV. The theoretical uncertainty on the scaling factor was evaluated by varying quark mass and the perturbative scales as described in ~\cite{scaling}. The D meson production cross sections were measured with ALICE, with limited precision and \pt\ coverage, in pp collisions at \sqs = 2.76 TeV~\cite{alice_charm_xsec}. These measurements were found to be in agreement with the scaled 7 TeV measurements within statistical uncertainties. A contribution from $\Lambda_\mathrm{c}$ decays was included using the measured ratio $\sigma(\Lambda_\mathrm{c})$/$\sigma(\mathrm{D^{0}+D^{+}})$ from ZEUS~\cite{zeus_lambdac}. The background electrons surviving the selection criteria, including the condition on \dzero, were subtracted from the measured electron distribution. Hadron contamination was estimated using a simultaneous fit of the electron and the different hadron components of the TPC d$E$/d$x$ distribution in momentum slices. The contamination was negligible below 4 GeV/$c$  but is significant at higher momenta. At 8 GeV/$c$ it was found to be approximately 7$\%$. The contamination was statistically subtracted from the measured electron distribution. The resulting \pt\ distribution is shown as filled circles in Fig. \ref{fig:rawpt}. 

The electron yield from beauty hadron decays was corrected for geometrical acceptance, track reconstruction efficiency, electron identification efficiency, and efficiency of the \dzero\ cut. The invariant cross section of inclusive electron production from beauty hadron decays in the range $|y| < 0.8$ was then calculated using the corrected electron \pt\ spectrum, the number of MB pp collisions and the MB cross section. The details are described in \cite{beauty7pp}.

\begin{table*}
\resizebox{\columnwidth}{!}{
\begin{tabular}{l l l}

\\
Uncertainty source & Systematic uncertainty (\%)\\ [0.6ex]

& 1$<$\pt$<$2 GeV/$c$ & 2$<$\pt$<$8 GeV/$c$  \\ [0.6ex]
\hline 
Track matching & $\pm$2  & $\pm$2 \\ 
ITS number of hits & $\pm$10  & $\pm$10 \\
Number of TPC clusters for tracking & +1,-10 &  $\pm$1 \\
Number of TPC clusters for PID &  $\pm$3 &  $\pm$3 \\
TOF  PID &  $\pm$3 &  n.a. \\
TPC PID &  $\pm$10 & $\pm$10 \\
Track $\eta$ and charge dependence &  $\pm$2  &  $\pm$2  \\
Minimum \dzero\ requirement & +15,-25 & $\pm$15 \\
Light hadron decay background & $\approx$15 & $<$ 3  \\
Charm hadron decay background & +40, -60 & $<$10  \\[1ex] 

\hline 
\end{tabular}
}
\caption{Contributions to the systematic uncertainty of the measurement of electrons from beauty hadron decays with the impact parameter method, for the ranges 1$<$\pt$<$2 GeV/$c$ (center column) and 2$<$\pt$<$8 GeV/$c$ (right column). The total systematic uncertainty is calculated as the quadrature sum of all contributions.}   
\label{table:sysimpact} 
\end{table*}

To evaluate systematic uncertainties, the analysis was repeated with modified track selection and Particle IDentification (PID) criteria. The contributions to the systematic uncertainty are listed in Table \ref{table:sysimpact}. The systematic uncertainties due to the tracking efficiencies and PID efficiencies are $^{+15}_{-18} (\pm 15)$\% for \mbox{\pt $<$ 2 GeV/$c$} (2 $<$ \pt  $<$ 6 GeV/$c$). These reach $\approx$$^{+20}_{-40}$\% at  8 GeV/$c$ due to the uncertainty of the hadron contamination subtraction, which is $\approx$$^{+8}_{-30}$\% at 8 GeV/$c$. Additional contributions to the total systematic uncertainty include the \dzero\ selection, evaluated by repeating the full analysis with modified selection criteria, and the subtraction of light flavor hadron decay background and charm hadron decay background, which were obtained by propagating the statistical and systematic uncertainties of the light flavor and charm hadron measurements used as analysis input. The light hadron decay background systematic uncertainty includes the uncertainty of the $m_{\mathrm{T}}$-scaling, which is conservatively taken to be 30$\%$. All systematic uncertainties were added in quadrature to obtain the total systematic uncertainty.

\subsection{Azimuthal electron-hadron correlation technique}

This analysis is based on the shape of the distribution of the difference in azimuth ($\Delta\varphi$) between electrons and hadrons, and in particular of the peak at $\Delta\varphi$ around zero (near-side). Due to the different decay kinematics of charm and beauty hadrons, the width of the near-side peak is larger for beauty than for charm hadron decays.  This method has been previously used by the STAR experiment \cite{star}. A similar method based on the the invariant mass of like charge sign electron-kaon pairs \cite{phenix} was used by the PHENIX experiment to extract a relative beauty contribution to the measured heavy-flavour electron production cross section.

\begin{figure}[tbh!]
\centering
\includegraphics[width=0.9\textwidth]{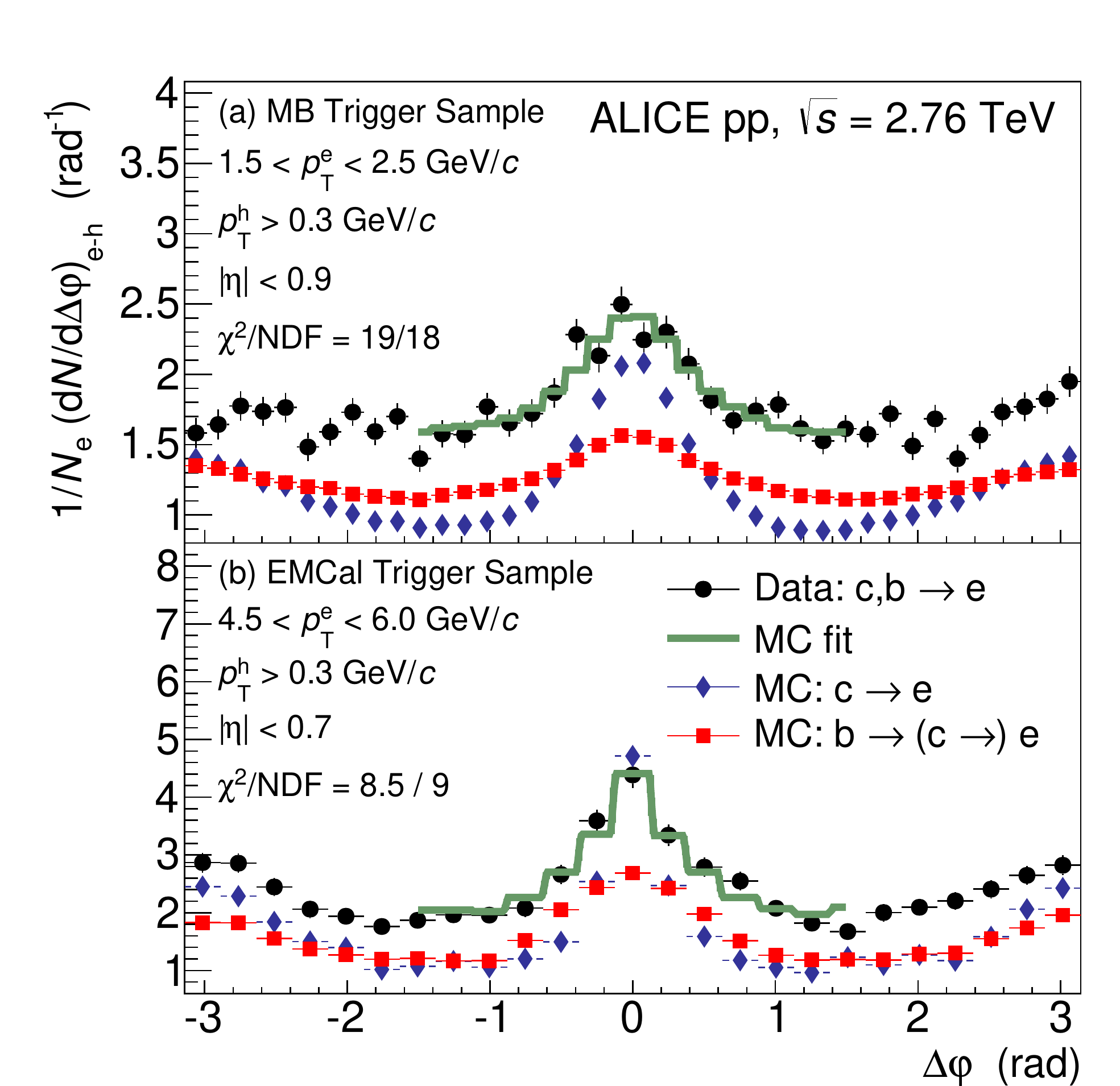}
\caption{(Color online) The azimuthal correlation between heavy-flavour decay electrons and charged hadrons, scaled by the number of electrons is shown for (a) the MB events in the $p^{\mathrm{e}}_{\mathrm{T}}$ range 1.5 to 2.5 GeV/$c$ and (b) the EMCal events in the $p^{\mathrm{e}}_{\mathrm{T}}$ range 4.5 to 6.0 GeV/$c$. The diamonds represent the MC distribution for electrons from charm hadron decays, squares are the MC distribution for electrons from beauty hadron decays. The line is the MC fit (Eq. \ref{eq:fit}) to the data points (circles).}
\label{fig:dPhiPtBins}       
\end{figure}

The analysis was performed using the MB and EMCal trigger data sets. Electrons were selected in the range 1 $< $ \pt\  $< $ 10 GeV/$c$. For the MB analysis the selected electrons reached out to a transverse momentum of 6 GeV/$c$, while the analysis using EMCal triggered events selects electrons in the range 2.5 $<$ \pt\ $<$ 10 GeV/$c$. 

The electron sample ($N_{\mathrm{e_{incl}}}$) contains electrons from heavy-flavour hadron decays and the aforementioned background sources, listed in Section 3.1. Di-electron pairs from photon conversions and $\pi^{0}$ Dalitz decays dominate at low \pt\ and were identified by pairing electrons with oppositely charged partner tracks and calculating the invariant mass ($M_{\mathrm{e}^+\mathrm{e}^-}$) of each e$^{+}$e$^{-}$ pair. The distribution for the background electrons is peaked at low $M_{\mathrm{e}^+\mathrm{e}^-}$, while no correlation signal is present in the low $M_{\mathrm{e}^+\mathrm{e}^-}$ region for the electrons from heavy-flavour decays. 
These unlike charge-sign (ULS) pairs contain true conversion and Dalitz decay electrons, along with a small fraction of heavy-flavour electrons that were wrongly paired with a background electron. The latter can be identified by calculating the invariant mass of like charge-sign (LS) pairs. Using a MC simulation with GEANT3, where pp collisions  are generated using PYTHIA 6 (Perugia-0 tune) and by comparing the ULS and LS invariant mass distribution the selection criteria on $M_{\mathrm{e}^+\mathrm{e}^-}$, identical for the LS and ULS pairs,  were determined. Electrons with $M_{\mathrm{e}^+\mathrm{e}^-}$ $<$ 50(100) MeV/$\mathrm{c^2}$ for the EMCal(MB) analysis were identified as background. The background finding efficiency ($\epsilon$) ranges from $\sim$ 20\% at low \pt\ to $\sim$ 66\% for \pt\ above 4 GeV/$c$. 

The number of heavy-flavour hadron decay electrons can be expressed as

\begin{equation}
N_{\mathrm{e_{HF}}} = N_{\mathrm{e_{incl}}} - \frac{1}{\epsilon} \left(N_{\mathrm{e_{ULS}}} -N_{\mathrm{e_{LS}}}\right),
\end{equation}

\noindent where $N_\mathrm{e_{ULS}}$ ($N_{\mathrm{e_{LS}}}$) are the number of electrons which formed a ULS(LS) pair with a $M_{\mathrm{e}^+\mathrm{e}^-}$ satisfying the previously mentioned selection criteria. Each electron contribution from Equation (1) is taken, along with the charged hadrons in the event and the heavy-flavour decay electron-hadron azimuthal correlation distribution, $\frac{1}{N_{\mathrm{e}}} \left( \frac{\mathrm{d}N}{\mathrm{d}\Delta\varphi}\right)_{\mathrm{e_{HF}-h}}$, was constructed.

To determine the fraction of electrons from beauty hadron decays the measured azimuthal e-h correlation distribution was fit with the function

 { \small
\begin{equation}
\hspace{-0.8cm}
\frac{1}{N_{\mathrm{e_{HF}}}} \left(\frac{{\mathrm d}N}{{\mathrm d}\Delta\varphi} \right)_{\mathrm{e_{HF}-h}}  = C + r_\mathrm{b} \frac{1}{N_{\mathrm{e_b}}} \left(\frac{{\mathrm d}N}{{\mathrm d}\Delta\varphi} \right)_{\mathrm{e_b-h}}  + (1-r_\mathrm{b})\frac{1}{N_{\mathrm{e}_{\mathrm{c}}}}\left(\frac{{\mathrm d}N}{{\mathrm d}\Delta\varphi} \right)_{\mathrm{e_c-h}}, 
\label{eq:fit}     
\end{equation}
}
\noindent where $r_b$, a free parameter of the fit, is the fraction of electrons from beauty to the total number of electrons from all heavy-flavour decays, $\Delta\varphi$ is the azimuthal angle between the electron and the charged hadron. The distributions of the azimuthal correlations $\left(\frac{\mathrm{d}N}{\mathrm{d}\Delta\varphi}\right)_{\mathrm{e_{b(c)}-h}}$ for electrons from beauty (charm) hadron decays were taken from the previously mentioned MC simulation, and the constant $C$ accounts for the uncorrelated background.  Fig. \ref{fig:dPhiPtBins} shows the measured azimuthal correlation, scaled by the number of electrons, along with the MC fit templates and the full fit for both (a) the MB and (b) the EMCal trigger analyses, in the \pt\ range of 1.5-2.5 GeV/$c$ and 4.5-6 GeV/$c$, respectively. For each \pt\ bin the measured distribution was fit on the near-side, over the range $|\Delta\varphi| < 1.5$ rad. From the fit, the relative beauty fraction ($r_b$) is extracted as a function of \pt. The values of $r_b$ obtained from the MB and EMCal triggered samples were found to agree within the systematic and statistical uncertainties in the overlapping \pt\ intervals. Hence, in the common \pt\ range, the final results for the relative beauty contribution to heavy-flavour decay electrons was obtained as the weighted average of the results from the MB and EMCal samples.

\begin{table}
\centering 
{
\begin{tabular}{l l l}
\\

Uncertainty source & Systematic uncertainty (\%) \\ [0.6ex]

& MB & EMCal \\ [0.6ex]
\hline 
Number of TPC clusters for tracking & $\pm$8 & 5  \\
TPC PID & $\pm$5(+5,-20) for \pt$<$($>$)3.5 GeV/$c$& $\pm$5($\pm$10) for \pt$<$($>$)3.5 GeV/$c$  \\
TOF PID & $\pm$5 & n.a.  \\
EMCal PID & n.a. & $\pm$10($\pm$5) for \pt$<$($>$)3.5 GeV/$c$  \\
$e^{+}e^{-}$ invariant mass & negligible & $\pm$10($\pm$5) for \pt$<$($>$)3.5 GeV/$c$ \\
Associated electron PID & $\pm$1& $\pm$1($\pm$5) for \pt$<$($>$)4.5 GeV/$c$ \\
Associated hadron momentum & $\pm$8 & $\pm$10($\pm$5) for \pt$<$($>$)3.5 GeV/$c$  \\ 
Fit range & negligible & negligible($\pm$5) for \pt$<$($>$6) GeV/$c$ \\
Light hadron decay background & $\pm$1 & $\pm$25($\pm$5) for \pt$<$($>$)3.5 GeV/$c$  \\[1ex] 
\hline 
\end{tabular}
}
\label{table:nonlin} 
\caption{Contributions to the systematic uncertainty of the fraction of electrons from beauty to the total number of electrons from heavy-flavour decays measured using the e-h azimuthal correlation technique, for the MB trigger (center column) and EMCal trigger (right column) analyses. The total systematic uncertainty is calculated as the quadrature sum of all contributions.} 
\end{table}

The main sources of systematic uncertainty include the electron identification selection criteria and the background finding efficiency. As previously explained, the background electrons were identified using invariant mass $M_{\mathrm{e}^+\mathrm{e}^-}$. The selected mass requirement, as a source of systematic uncertainty was found to be negligible for the MB analysis and reached a maximum of 10\% for \pt\ $<$ 3.5 GeV for the EMCal analysis. The efficiency of the invariant mass method was calculated using a MC sample.  For the EMCal analysis a MC simulation enhanced with $\pi^{0}$ and $\eta$ mesons, flat in \pt, was used in order to increase statistics of background electrons at high \pt, as the MB MC sample did not provide enough statistics. The bias from the enhancement is corrected by reweighting to obtain the correct \pt\ -distribution of the $\pi^{0}$ (see Section 3.1). Overall, the systematic uncertainties range from 9 to 21\% for the MB analysis and from 12 to 33\% in the case of the EMCal analysis, depending on the transverse momentum. The final systematic uncertainties were obtained by combining these two measurements, yielding 17\% for the lower momentum region (\pt\ $<$ 3.5 GeV/$c$) and $^{+16}_{-25}$\% for the higher momentum region (3.5 $<$  \pt\ $<$ 10 GeV/$c$). All systematic uncertainties are listed in Table \ref{table:nonlin}. 

For the MB analysis the hadron contamination to the electron sample was estimated using a simultaneous fit of the electron and the different hadron components of the TPC dE/dx distribution in momentum ranges, while for the EMCal analysis the contamination  was estimated using a fit to the $E/p$ distribution in momentum slices. The contamination was found to be negligible for \pt\ $ < $ 4(6) GeV/$c$ for the MB(EMCal) analysis. For the highest \pt\ of the MB analysis the contamination was 5$\%$ and reached ~20$\%$ for the highest \pt\ of the EMCal analysis. No subtraction of this contamination was performed. Instead it is taken into account in the PID systematic uncertainties. In addition, a mixed event technique was used to cross-check that detector acceptance effects are well reproduced in the MC sample. For the mixed event $\Delta\varphi$ correlation distribution, electrons from EMCal trigger events and hadrons from the MB sample were selected. Hadrons were selected only from MB events to remove the bias from EMCal trigger sample in the correlation distribution from mixed event. The mixed event correlation distribution was found to be flat over the entire $\Delta\varphi$ range, implying that detector effects do not bias the correlation distribution. Hence a mixed event correction was not applied to the resulting $\Delta\varphi$ distribution. 

\section{Results}
\label{results}

\begin{figure}[tbh!]
\centering
\includegraphics[width=0.9\textwidth]{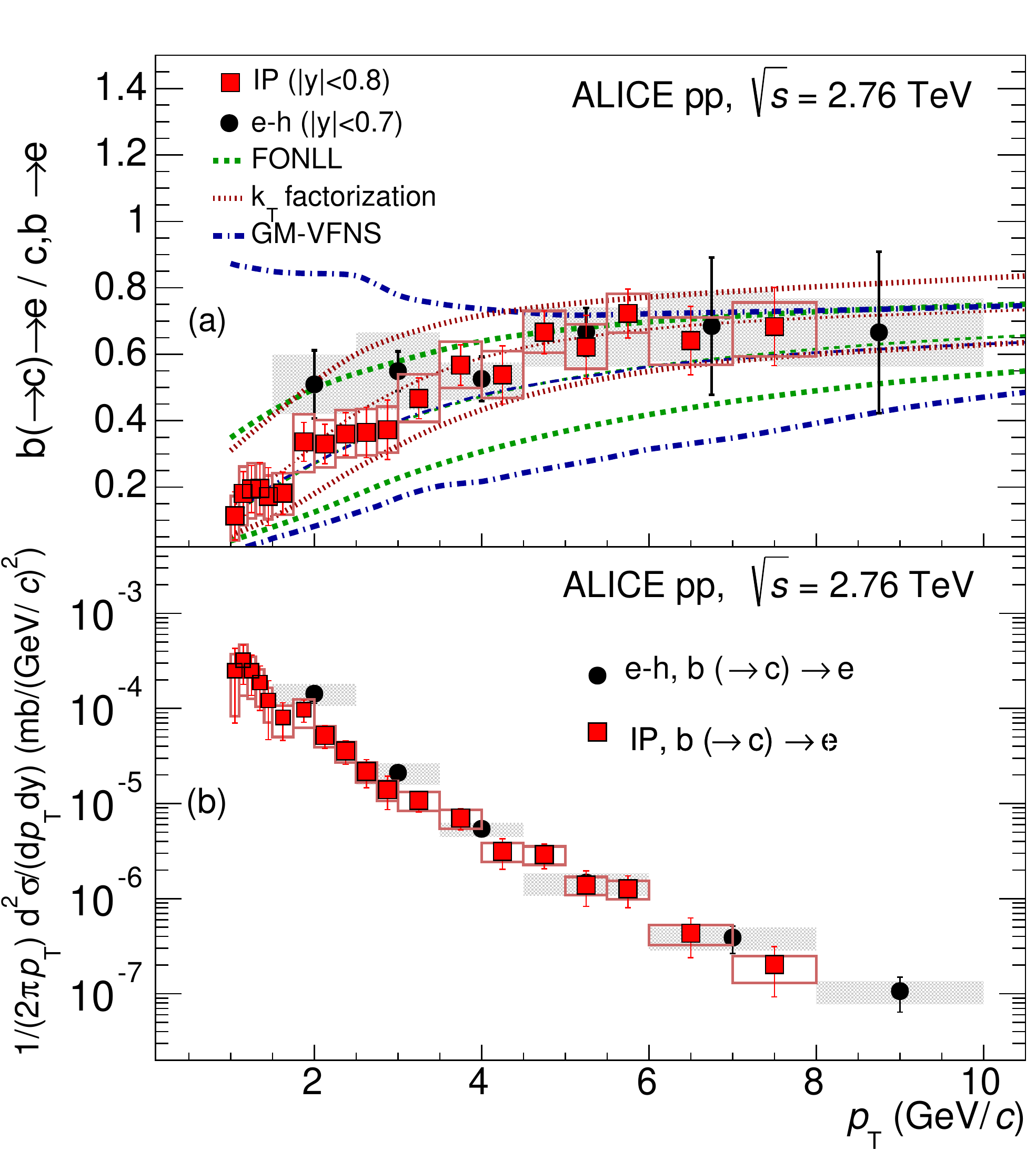}
\caption{(Color online) (a) Relative beauty contribution to the heavy-flavour electron yield; measured from the azimuthal correlations between heavy-flavour decay electrons and charged hadrons (black circles) compared to that from the method based on the track impact parameter (red squares). The green dashed, red dotted, and blue dot-dashed lines represent the FONLL \cite{fonll3}, $k_{\rm T}$-factorization \cite{Maciula:2013wg}, and GM-VFNS \cite{Bolzoni:2012kx} predictions, respectively. (b) The \pt -differential inclusive production cross section of electrons from beauty hadron decays obtained using the impact parameter method (red squares) and the e-h correlation (black circles) method. For both panels, the error bars (boxes) represent the statistical (systematic) uncertainties. The notation $\rm b (\rightarrow \rm c) \rightarrow \rm e$ is used to indicate that the relative beauty contribution includes those electrons which originate directly from beauty hadron decays and those which originate from charm hadron decays, where the charm hadron is the decay product of a beauty hadron.}
\label{fig:bfraction}       
\end{figure}

The relative beauty contribution to heavy-flavour decay electrons obtained from the impact parameter analysis, along with that extracted from the azimuthal correlation method, is shown as a function of \pt\ in Fig. \ref{fig:bfraction}(a). For the impact parameter analysis the beauty contribution to the heavy-flavour electron spectrum was measured, while the charm contribution was calculated from the charm hadron spectra measured by ALICE as described in Section 3.1. Within the statistical and systematic uncertainties the resulting fractions are in agreement with each other and show that the beauty contribution to the total heavy-flavour spectrum is comparable to the contribution from charm for \pt\ $>$ 4 GeV/$c$. 

\begin{figure}[tbh!]
\centering
\includegraphics[width=0.9\textwidth]{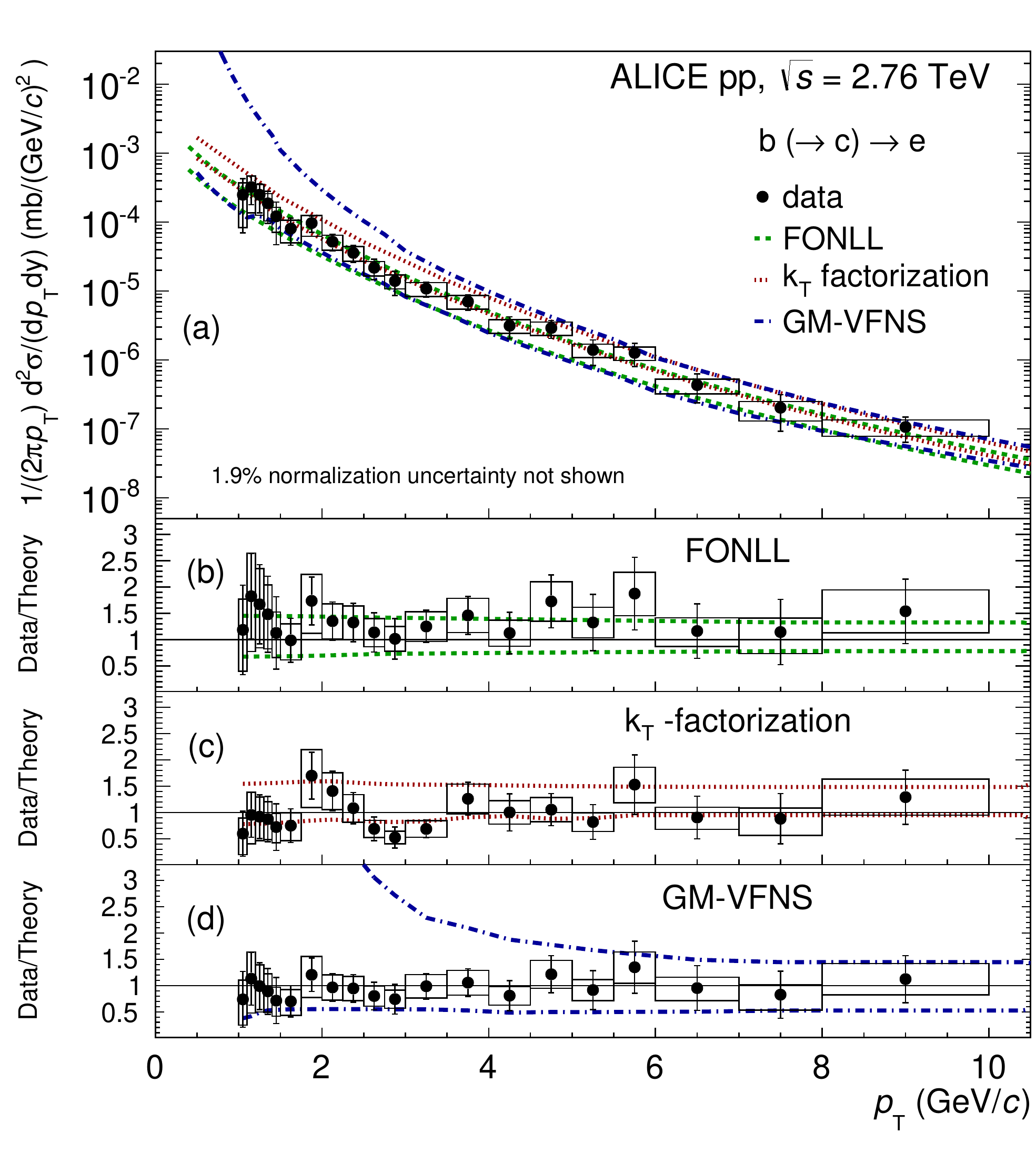}
\caption{(Color online) (a) \pt -differential inclusive production cross section of electrons from beauty hadron decays. The green dashed, red dotted, and blue dot-dashed lines represent the FONLL \cite{fonll3}, $k_{\rm T}$-factorization \cite{Maciula:2013wg}, and GM-VFNS \cite{Bolzoni:2012kx} uncertainty range, respectively. (b)-(d) Ratios of the data and the central prediction of pQCD calculations for electrons from beauty  hadron decays. For all panels, the error bars (boxes) represent the statistical (systematic) uncertainties}
\label{fig:btoe}       
\end{figure}

The measurements are compared to the central, upper, and lower predictions of three sets of pQCD calculations \cite{fonll3,Bolzoni:2012kx,Maciula:2013wg}, represented by the various lines. The central values of the fraction of electrons from beauty hadron decays were calculated using the central values of the beauty and charm to electron cross sections. The upper (lower) predictions were obtained by calculating the beauty fraction using the upper (lower) uncertainty limit of the beauty to electron cross section and the lower (upper) limit of the charm to electron cross section. The upper and lower lines demonstrate the uncertainty range of the calculations, which originate from the variation of the perturbative scales and the heavy quark masses as described in \cite{fonll3,gmvfns,Maciula:2013wg}. Each prediction describes the relative beauty contribution fraction over the whole \pt\ range.

The \pt -differential production cross section of electrons from beauty hadron decays measured using the impact parameter analysis is shown in Fig. \ref{fig:bfraction} (b) and it is compared to the spectrum obtained using the beauty fraction from the e-h correlation analysis and the measured heavy-flavour decay electron cross section from \cite{alice_hfe}. This alternative approach agrees with the result obtained using the impact parameter technique. As the resulting spectrum obtained using the impact parameter based analysis ($|y| < 0.8$) yielded finer \pt\ intervals and smaller uncertainties this result for \pt\ $<$ 8 GeV/$c$ is used with the higher \pt\ slice of the e-h correlation analysis ($|y| < 0.7$) to obtain the total beauty production cross section. 

The measured \pt -differential cross section, obtained using the impact parameter analysis for \pt\ $<$ 8 GeV/$c$ and including the highest \pt\ point from the correlation analysis, in the \pt\ range 1-10 GeV/$c$ is shown in Fig. \ref{fig:btoe} (a) along with a comparison to the upper and lower uncertainty limits  of the aforementioned pQCD calculations.  Fig. \ref{fig:btoe} (b)-(d) shows the ratio of the data to the central theoretical predictions. The data and predictions are consistent within the experimental and theoretical uncertainties. Due to the uncertainty of the measured luminosity all measured cross sections have an additional normalization uncertainty of 1.9\% \cite{alice_cross_section}.

 \begin{figure}[tbh!]
\centering
\includegraphics[width=0.9\textwidth]{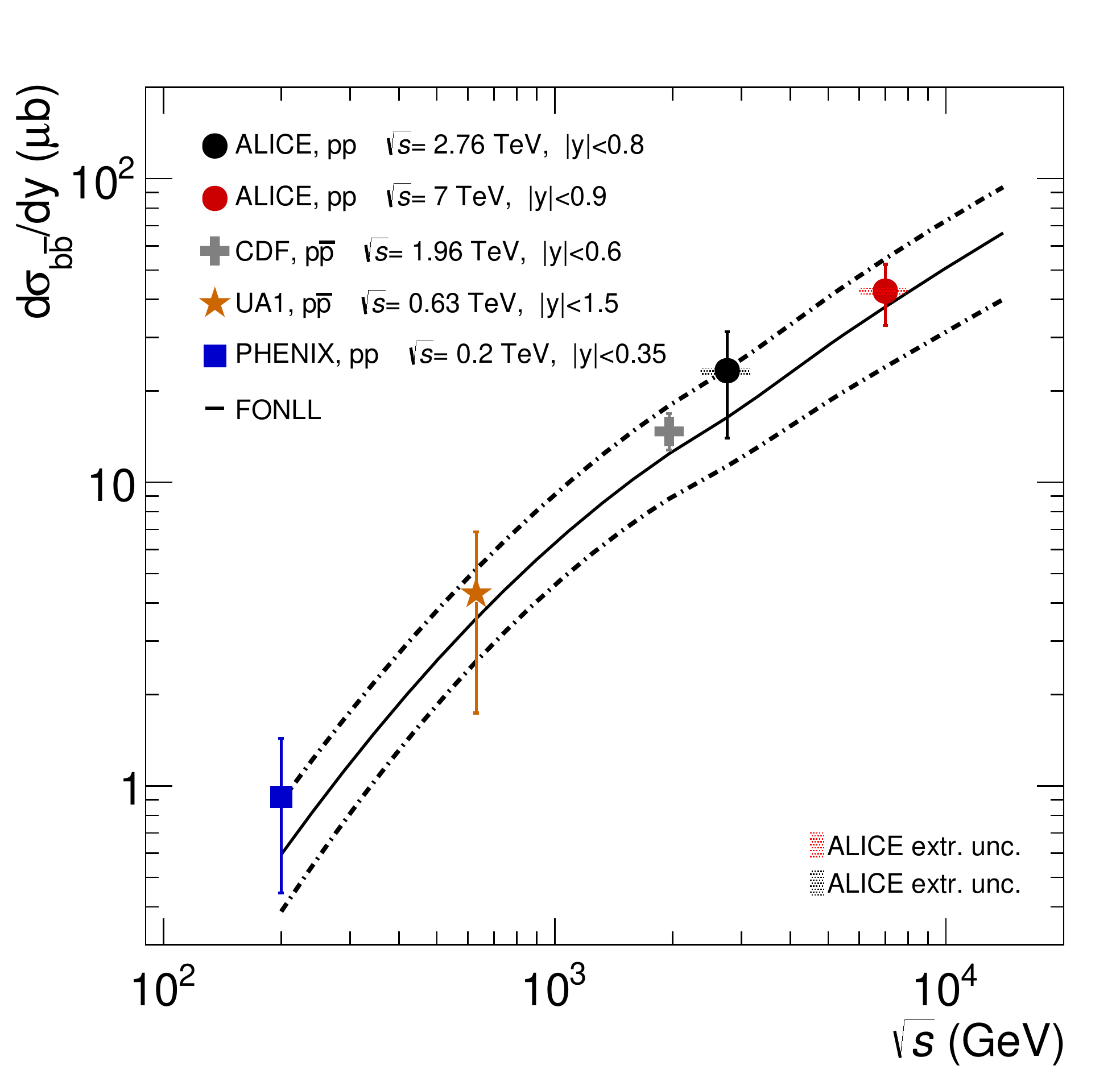}
\caption{(Color online) Inclusive beauty production cross section per rapidity unit measured at mid-rapidity as a function of center of mass energy in pp collisions (PHENIX \cite{phenix} and ALICE \cite{beauty7pp} results) and p$\bar{\mathrm{p}}$ collisions (UA1 \cite{uaone} and CDF \cite{cdf} results) along with the the comparison to FONLL calculations. Error bars represent the statistical and systematic uncertainties added in quadrature. The FONLL calculation was performed for the five experimental rapidity ranges and center of mass energies shown in the figure, and these points are drawn as a curve.}
\label{fig:totalbbxsec}       
\end{figure}
 
The visible cross section of electrons from beauty hadron decays at mid-rapidity ($|y|<0.8$) was obtained by integrating the \pt -differential cross section in the measured \pt\ range ($1 < p_{\mathrm{T}}<10$ GeV/$c$), obtaining $\sigma_{\mathrm{b} \rightarrow \mathrm{e}} = 3.47\pm0.40(\mathrm{stat})^{+1.12}_{-1.33}(\mathrm{sys})\pm0.07(\mathrm{norm})~ \mu$b. The visible cross section is then scaled by the ratio of the total cross section of electrons originating from beauty hadron decays from FONLL in the full \pt\ range to the FONLL cross section integrated in the measured \pt\ range. The central value of the extrapolation factor was computed using the FONLL prediction with the central values of the quark mass and perturbative scale. The uncertainties were obtained by varying the quark mass and perturbative scale and recalculating the ratio, which is given separately in the results as extrapolation uncertainty. For the extrapolation the beauty hadron to electron branching ratio of BR$_{\mathrm{H_b\rightarrow e}} + \mathrm{BR}_{\mathrm{H_b\rightarrow H_c \rightarrow e}} = 0.205\pm0.007$ \cite{pdg} is used.

The beauty production cross section at mid-rapidity, per unit rapidity, $\frac{\mathrm{d}\sigma_{\mathrm{b\bar{b}}}}{\mathrm{d}y} = 23.28\pm2.70(\mathrm{stat})^{+8.92}_{-8.70}\\*(\mathrm{sys})^{+0.49}_{-0.65}(\mathrm{extr})\pm0.44(\mathrm{norm})~ \mu$b, is shown in Fig. \ref{fig:totalbbxsec} as a function of center of mass energy for experimental measurements \cite{phenix,cdf, uaone}, including the result obtained by ALICE at 7 TeV \cite{beauty7pp}. The total beauty production cross section was obtained by extrapolating to the full $y$ range and is found to be $\sigma_{\mathrm{b\bar{b}}} = 130\pm15.1(\mathrm{stat})^{+42.1}_{-49.8}(\mathrm{sys})^{+3.4}_{-3.1}(\mathrm{extr})\pm2.5(\mathrm{norm})\pm4.4(\mathrm{BR})~ \mu$b. The corresponding prediction of FONLL is $\sigma_{\mathrm{b\bar{b}}} = 95.5^{+139}_{-66.5}~ \mu $b. 

\section{Summary}
\label{summary}

The inclusive invariant production cross section of electrons from semi-leptonic decays of beauty hadrons is reported at mid-rapidity ($|y| < 0.8$) in the transverse momentum range 1 $<$ \pt\ $<$ 10 GeV/$c$, in pp collisions at \sqs = 2.76 TeV. The primary measurement utilized a selection of tracks based on their impact parameter to identify displaced electrons from beauty hadron decays. An alternative method, which utilized the measured electron-hadron azimuthal correlations, was found to be in agreement with the results from the impact parameter method. The results are compared to pQCD calculations and agreement between data and theory was found. The integrated visible cross section is $\sigma_{\mathrm{b} \rightarrow \mathrm{e}} = 3.47\pm0.40(\mathrm{stat})^{+1.12}_{-1.33}(\mathrm{sys})\pm0.07(\mathrm{norm})~ \mu$b. Extrapolation to full phase space using FONLL yields the total b$\bar{\mathrm{b}}$ production cross section, $\sigma_{\mathrm{b\bar{b}}} = 130\pm15.1(\mathrm{stat})^{+42.1}_{-49.8}(\mathrm{sys})^{+3.4}_{-3.1}(\mathrm{extr})\pm2.5(\mathrm{norm})\pm4.4(\mathrm{BR})~ \mu$b. These results provide a crucial reference for the study of beauty quark production in Pb-Pb collisions at the LHC.

\newenvironment{acknowledgement}{\relax}{\relax}
\begin{acknowledgement}
\section*{Acknowledgements}
The ALICE Collaboration would like to thank all its engineers and technicians for their invaluable contributions to the construction of the experiment and the CERN accelerator teams for the outstanding performance of the LHC complex.
The ALICE Collaboration gratefully acknowledges the resources and support provided by all Grid centres and the Worldwide LHC Computing Grid (WLCG) collaboration.
The ALICE Collaboration acknowledges the following funding agencies for their support in building and
running the ALICE detector:
State Committee of Science,  World Federation of Scientists (WFS)
and Swiss Fonds Kidagan, Armenia,
Conselho Nacional de Desenvolvimento Cient\'{\i}fico e Tecnol\'{o}gico (CNPq), Financiadora de Estudos e Projetos (FINEP),
Funda\c{c}\~{a}o de Amparo \`{a} Pesquisa do Estado de S\~{a}o Paulo (FAPESP);
National Natural Science Foundation of China (NSFC), the Chinese Ministry of Education (CMOE)
and the Ministry of Science and Technology of China (MSTC);
Ministry of Education and Youth of the Czech Republic;
Danish Natural Science Research Council, the Carlsberg Foundation and the Danish National Research Foundation;
The European Research Council under the European Community's Seventh Framework Programme;
Helsinki Institute of Physics and the Academy of Finland;
French CNRS-IN2P3, the `Region Pays de Loire', `Region Alsace', `Region Auvergne' and CEA, France;
German BMBF and the Helmholtz Association;
General Secretariat for Research and Technology, Ministry of
Development, Greece;
Hungarian OTKA and National Office for Research and Technology (NKTH);
Department of Atomic Energy and Department of Science and Technology of the Government of India;
Istituto Nazionale di Fisica Nucleare (INFN) and Centro Fermi -
Museo Storico della Fisica e Centro Studi e Ricerche "Enrico
Fermi", Italy;
MEXT Grant-in-Aid for Specially Promoted Research, Ja\-pan;
Joint Institute for Nuclear Research, Dubna;
National Research Foundation of Korea (NRF);
CONACYT, DGAPA, M\'{e}xico, ALFA-EC and the EPLANET Program
(European Particle Physics Latin American Network)
Stichting voor Fundamenteel Onderzoek der Materie (FOM) and the Nederlandse Organisatie voor Wetenschappelijk Onderzoek (NWO), Netherlands;
Research Council of Norway (NFR);
Polish Ministry of Science and Higher Education;
National Science Centre, Poland;
 Ministry of National Education/Institute for Atomic Physics and CNCS-UEFISCDI - Romania;
Ministry of Education and Science of Russian Federation, Russian
Academy of Sciences, Russian Federal Agency of Atomic Energy,
Russian Federal Agency for Science and Innovations and The Russian
Foundation for Basic Research;
Ministry of Education of Slovakia;
Department of Science and Technology, South Africa;
CIEMAT, EELA, Ministerio de Econom\'{i}a y Competitividad (MINECO) of Spain, Xunta de Galicia (Conseller\'{\i}a de Educaci\'{o}n),
CEA\-DEN, Cubaenerg\'{\i}a, Cuba, and IAEA (International Atomic Energy Agency);
Swedish Research Council (VR) and Knut $\&$ Alice Wallenberg
Foundation (KAW);
Ukraine Ministry of Education and Science;
United Kingdom Science and Technology Facilities Council (STFC);
The United States Department of Energy, the United States National
Science Foundation, the State of Texas, and the State of Ohio.
    
\end{acknowledgement}

\bibliographystyle{utphys}   
\bibliography{hfe_pp_beauty}

\newpage

\appendix
\section{The ALICE Collaboration}
\label{app:collab}



\begingroup
\small
\begin{flushleft}
B.~Abelev\Irefn{org71}\And
J.~Adam\Irefn{org37}\And
D.~Adamov\'{a}\Irefn{org79}\And
M.M.~Aggarwal\Irefn{org83}\And
G.~Aglieri~Rinella\Irefn{org34}\And
M.~Agnello\Irefn{org107}\textsuperscript{,}\Irefn{org90}\And
A.~Agostinelli\Irefn{org26}\And
N.~Agrawal\Irefn{org44}\And
Z.~Ahammed\Irefn{org126}\And
N.~Ahmad\Irefn{org18}
% ...
\renewcommand\labelenumi{\textsuperscript{\theenumi}~}

\section*{Affiliation notes}
\renewcommand\theenumi{\roman{enumi}}
\begin{Authlist}
\item \Adef{0}Deceased
\item \Adef{idp1118016}{Also at: St. Petersburg State Polytechnical University}
\item \Adef{idp3027568}{Also at: Department of Applied Physics, Aligarh Muslim University, Aligarh, India}
\item \Adef{idp3708288}{Also at: M.V. Lomonosov Moscow State University, D.V. Skobeltsyn Institute of Nuclear Physics, Moscow, Russia}
\item \Adef{idp3951888}{Also at: University of Belgrade, Faculty of Physics and "Vin\v{c}a" Institute of Nuclear Sciences, Belgrade, Serbia}
\item \Adef{idp4263616}{Permanent Address: Permanent Address: Konkuk University, Seoul, Korea}
\item \Adef{idp4814880}{Also at: Institute of Theoretical Physics, University of Wroclaw, Wroclaw, Poland}
\item \Adef{idp5735392}{Also at: University of Kansas, Lawrence, KS, United States}
\end{Authlist}

\section*{Collaboration Institutes}
\renewcommand\theenumi{\arabic{enumi}~}
\begin{Authlist}

\item \Idef{org1}A.I. Alikhanyan National Science Laboratory (Yerevan Physics Institute) Foundation, Yerevan, Armenia
\item \Idef{org2}Benem\'{e}rita Universidad Aut\'{o}noma de Puebla, Puebla, Mexico
\item \Idef{org3}Bogolyubov Institute for Theoretical Physics, Kiev, Ukraine
\item \Idef{org4}Bose Institute, Department of Physics and Centre for Astroparticle Physics and Space Science (CAPSS), Kolkata, India
\item \Idef{org5}Budker Institute for Nuclear Physics, Novosibirsk, Russia
\item \Idef{org6}California Polytechnic State University, San Luis Obispo, CA, United States
\item \Idef{org7}Central China Normal University, Wuhan, China
\item \Idef{org8}Centre de Calcul de l'IN2P3, Villeurbanne, France
\item \Idef{org9}Centro de Aplicaciones Tecnol\'{o}gicas y Desarrollo Nuclear (CEADEN), Havana, Cuba
\item \Idef{org10}Centro de Investigaciones Energ\'{e}ticas Medioambientales y Tecnol\'{o}gicas (CIEMAT), Madrid, Spain
\item \Idef{org11}Centro de Investigaci\'{o}n y de Estudios Avanzados (CINVESTAV), Mexico City and M\'{e}rida, Mexico
\item \Idef{org12}Centro Fermi - Museo Storico della Fisica e Centro Studi e Ricerche ``Enrico Fermi'', Rome, Italy
\item \Idef{org13}Chicago State University, Chicago, USA
\item \Idef{org14}Commissariat \`{a} l'Energie Atomique, IRFU, Saclay, France
\item \Idef{org15}COMSATS Institute of Information Technology (CIIT), Islamabad, Pakistan
\item \Idef{org16}Departamento de F\'{\i}sica de Part\'{\i}culas and IGFAE, Universidad de Santiago de Compostela, Santiago de Compostela, Spain
\item \Idef{org17}Department of Physics and Technology, University of Bergen, Bergen, Norway
\item \Idef{org18}Department of Physics, Aligarh Muslim University, Aligarh, India
\item \Idef{org19}Department of Physics, Ohio State University, Columbus, OH, United States
\item \Idef{org20}Department of Physics, Sejong University, Seoul, South Korea
\item \Idef{org21}Department of Physics, University of Oslo, Oslo, Norway
\item \Idef{org22}Dipartimento di Fisica dell'Universit\`{a} 'La Sapienza' and Sezione INFN Rome, Italy
\item \Idef{org23}Dipartimento di Fisica dell'Universit\`{a} and Sezione INFN, Cagliari, Italy
\item \Idef{org24}Dipartimento di Fisica dell'Universit\`{a} and Sezione INFN, Trieste, Italy
\item \Idef{org25}Dipartimento di Fisica dell'Universit\`{a} and Sezione INFN, Turin, Italy
\item \Idef{org26}Dipartimento di Fisica e Astronomia dell'Universit\`{a} and Sezione INFN, Bologna, Italy
\item \Idef{org27}Dipartimento di Fisica e Astronomia dell'Universit\`{a} and Sezione INFN, Catania, Italy
\item \Idef{org28}Dipartimento di Fisica e Astronomia dell'Universit\`{a} and Sezione INFN, Padova, Italy
\item \Idef{org29}Dipartimento di Fisica `E.R.~Caianiello' dell'Universit\`{a} and Gruppo Collegato INFN, Salerno, Italy
\item \Idef{org30}Dipartimento di Scienze e Innovazione Tecnologica dell'Universit\`{a} del  Piemonte Orientale and Gruppo Collegato INFN, Alessandria, Italy
\item \Idef{org31}Dipartimento Interateneo di Fisica `M.~Merlin' and Sezione INFN, Bari, Italy
\item \Idef{org32}Division of Experimental High Energy Physics, University of Lund, Lund, Sweden
\item \Idef{org33}Eberhard Karls Universit\"{a}t T\"{u}bingen, T\"{u}bingen, Germany
\item \Idef{org34}European Organization for Nuclear Research (CERN), Geneva, Switzerland
\item \Idef{org35}Faculty of Engineering, Bergen University College, Bergen, Norway
\item \Idef{org36}Faculty of Mathematics, Physics and Informatics, Comenius University, Bratislava, Slovakia
\item \Idef{org37}Faculty of Nuclear Sciences and Physical Engineering, Czech Technical University in Prague, Prague, Czech Republic
\item \Idef{org38}Faculty of Science, P.J.~\v{S}af\'{a}rik University, Ko\v{s}ice, Slovakia
\item \Idef{org39}Frankfurt Institute for Advanced Studies, Johann Wolfgang Goethe-Universit\"{a}t Frankfurt, Frankfurt, Germany
\item \Idef{org40}Gangneung-Wonju National University, Gangneung, South Korea
\item \Idef{org41}Gauhati University, Department of Physics, Guwahati, India
\item \Idef{org42}Helsinki Institute of Physics (HIP), Helsinki, Finland
\item \Idef{org43}Hiroshima University, Hiroshima, Japan
\item \Idef{org44}Indian Institute of Technology Bombay (IIT), Mumbai, India
\item \Idef{org45}Indian Institute of Technology Indore, Indore (IITI), India
\item \Idef{org46}Inha University, Incheon, South Korea
\item \Idef{org47}Institut de Physique Nucl\'eaire d'Orsay (IPNO), Universit\'e Paris-Sud, CNRS-IN2P3, Orsay, France
\item \Idef{org48}Institut f\"{u}r Informatik, Johann Wolfgang Goethe-Universit\"{a}t Frankfurt, Frankfurt, Germany
\item \Idef{org49}Institut f\"{u}r Kernphysik, Johann Wolfgang Goethe-Universit\"{a}t Frankfurt, Frankfurt, Germany
\item \Idef{org50}Institut f\"{u}r Kernphysik, Westf\"{a}lische Wilhelms-Universit\"{a}t M\"{u}nster, M\"{u}nster, Germany
\item \Idef{org51}Institut Pluridisciplinaire Hubert Curien (IPHC), Universit\'{e} de Strasbourg, CNRS-IN2P3, Strasbourg, France
\item \Idef{org52}Institute for Nuclear Research, Academy of Sciences, Moscow, Russia
\item \Idef{org53}Institute for Subatomic Physics of Utrecht University, Utrecht, Netherlands
\item \Idef{org54}Institute for Theoretical and Experimental Physics, Moscow, Russia
\item \Idef{org55}Institute of Experimental Physics, Slovak Academy of Sciences, Ko\v{s}ice, Slovakia
\item \Idef{org56}Institute of Physics, Academy of Sciences of the Czech Republic, Prague, Czech Republic
\item \Idef{org57}Institute of Physics, Bhubaneswar, India
\item \Idef{org58}Institute of Space Science (ISS), Bucharest, Romania
\item \Idef{org59}Instituto de Ciencias Nucleares, Universidad Nacional Aut\'{o}noma de M\'{e}xico, Mexico City, Mexico
\item \Idef{org60}Instituto de F\'{\i}sica, Universidad Nacional Aut\'{o}noma de M\'{e}xico, Mexico City, Mexico
\item \Idef{org61}iThemba LABS, National Research Foundation, Somerset West, South Africa
\item \Idef{org62}Joint Institute for Nuclear Research (JINR), Dubna, Russia
\item \Idef{org63}Konkuk University, Seoul, South Korea
\item \Idef{org64}Korea Institute of Science and Technology Information, Daejeon, South Korea
\item \Idef{org65}KTO Karatay University, Konya, Turkey
\item \Idef{org66}Laboratoire de Physique Corpusculaire (LPC), Clermont Universit\'{e}, Universit\'{e} Blaise Pascal, CNRS--IN2P3, Clermont-Ferrand, France
\item \Idef{org67}Laboratoire de Physique Subatomique et de Cosmologie, Universit\'{e} Grenoble-Alpes, CNRS-IN2P3, Grenoble, France
\item \Idef{org68}Laboratori Nazionali di Frascati, INFN, Frascati, Italy
\item \Idef{org69}Laboratori Nazionali di Legnaro, INFN, Legnaro, Italy
\item \Idef{org70}Lawrence Berkeley National Laboratory, Berkeley, CA, United States
\item \Idef{org71}Lawrence Livermore National Laboratory, Livermore, CA, United States
\item \Idef{org72}Moscow Engineering Physics Institute, Moscow, Russia
\item \Idef{org73}National Centre for Nuclear Studies, Warsaw, Poland
\item \Idef{org74}National Institute for Physics and Nuclear Engineering, Bucharest, Romania
\item \Idef{org75}National Institute of Science Education and Research, Bhubaneswar, India
\item \Idef{org76}Niels Bohr Institute, University of Copenhagen, Copenhagen, Denmark
\item \Idef{org77}Nikhef, National Institute for Subatomic Physics, Amsterdam, Netherlands
\item \Idef{org78}Nuclear Physics Group, STFC Daresbury Laboratory, Daresbury, United Kingdom
\item \Idef{org79}Nuclear Physics Institute, Academy of Sciences of the Czech Republic, \v{R}e\v{z} u Prahy, Czech Republic
\item \Idef{org80}Oak Ridge National Laboratory, Oak Ridge, TN, United States
\item \Idef{org81}Petersburg Nuclear Physics Institute, Gatchina, Russia
\item \Idef{org82}Physics Department, Creighton University, Omaha, NE, United States
\item \Idef{org83}Physics Department, Panjab University, Chandigarh, India
\item \Idef{org84}Physics Department, University of Athens, Athens, Greece
\item \Idef{org85}Physics Department, University of Cape Town, Cape Town, South Africa
\item \Idef{org86}Physics Department, University of Jammu, Jammu, India
\item \Idef{org87}Physics Department, University of Rajasthan, Jaipur, India
\item \Idef{org88}Physik Department, Technische Universit\"{a}t M\"{u}nchen, Munich, Germany
\item \Idef{org89}Physikalisches Institut, Ruprecht-Karls-Universit\"{a}t Heidelberg, Heidelberg, Germany
\item \Idef{org90}Politecnico di Torino, Turin, Italy
\item \Idef{org91}Purdue University, West Lafayette, IN, United States
\item \Idef{org92}Pusan National University, Pusan, South Korea
\item \Idef{org93}Research Division and ExtreMe Matter Institute EMMI, GSI Helmholtzzentrum f\"ur Schwerionenforschung, Darmstadt, Germany
\item \Idef{org94}Rudjer Bo\v{s}kovi\'{c} Institute, Zagreb, Croatia
\item \Idef{org95}Russian Federal Nuclear Center (VNIIEF), Sarov, Russia
\item \Idef{org96}Russian Research Centre Kurchatov Institute, Moscow, Russia
\item \Idef{org97}Saha Institute of Nuclear Physics, Kolkata, India
\item \Idef{org98}School of Physics and Astronomy, University of Birmingham, Birmingham, United Kingdom
\item \Idef{org99}Secci\'{o}n F\'{\i}sica, Departamento de Ciencias, Pontificia Universidad Cat\'{o}lica del Per\'{u}, Lima, Peru
\item \Idef{org100}Sezione INFN, Bari, Italy
\item \Idef{org101}Sezione INFN, Bologna, Italy
\item \Idef{org102}Sezione INFN, Cagliari, Italy
\item \Idef{org103}Sezione INFN, Catania, Italy
\item \Idef{org104}Sezione INFN, Padova, Italy
\item \Idef{org105}Sezione INFN, Rome, Italy
\item \Idef{org106}Sezione INFN, Trieste, Italy
\item \Idef{org107}Sezione INFN, Turin, Italy
\item \Idef{org108}SSC IHEP of NRC Kurchatov institute, Protvino, Russia
\item \Idef{org109}SUBATECH, Ecole des Mines de Nantes, Universit\'{e} de Nantes, CNRS-IN2P3, Nantes, France
\item \Idef{org110}Suranaree University of Technology, Nakhon Ratchasima, Thailand
\item \Idef{org111}Technical University of Split FESB, Split, Croatia
\item \Idef{org112}The Henryk Niewodniczanski Institute of Nuclear Physics, Polish Academy of Sciences, Cracow, Poland
\item \Idef{org113}The University of Texas at Austin, Physics Department, Austin, TX, USA
\item \Idef{org114}Universidad Aut\'{o}noma de Sinaloa, Culiac\'{a}n, Mexico
\item \Idef{org115}Universidade de S\~{a}o Paulo (USP), S\~{a}o Paulo, Brazil
\item \Idef{org116}Universidade Estadual de Campinas (UNICAMP), Campinas, Brazil
\item \Idef{org117}University of Houston, Houston, TX, United States
\item \Idef{org118}University of Jyv\"{a}skyl\"{a}, Jyv\"{a}skyl\"{a}, Finland
\item \Idef{org119}University of Liverpool, Liverpool, United Kingdom
\item \Idef{org120}University of Tennessee, Knoxville, TN, United States
\item \Idef{org121}University of Tokyo, Tokyo, Japan
\item \Idef{org122}University of Tsukuba, Tsukuba, Japan
\item \Idef{org123}University of Zagreb, Zagreb, Croatia
\item \Idef{org124}Universit\'{e} de Lyon, Universit\'{e} Lyon 1, CNRS/IN2P3, IPN-Lyon, Villeurbanne, France
\item \Idef{org125}V.~Fock Institute for Physics, St. Petersburg State University, St. Petersburg, Russia
\item \Idef{org126}Variable Energy Cyclotron Centre, Kolkata, India
\item \Idef{org127}Vestfold University College, Tonsberg, Norway
\item \Idef{org128}Warsaw University of Technology, Warsaw, Poland
\item \Idef{org129}Wayne State University, Detroit, MI, United States
\item \Idef{org130}Wigner Research Centre for Physics, Hungarian Academy of Sciences, Budapest, Hungary
\item \Idef{org131}Yale University, New Haven, CT, United States
\item \Idef{org132}Yonsei University, Seoul, South Korea
\item \Idef{org133}Zentrum f\"{u}r Technologietransfer und Telekommunikation (ZTT), Fachhochschule Worms, Worms, Germany
\end{Authlist}
\endgroup

\end{document}